# EMPIRICAL FOUNDATIONS OF THE RELATIVISTIC GRAVITY

WEI-TOU NI

*Center for Gravitation and Cosmology, Solar-System Division,*
*Purple Mountain Observatory, Chinese Academy of Sciences,*
*No. 2, Beijing W. Rd., Nanjing, China 210008* wtni@pmo.ac.cn

In 1859, Le Verrier discovered the mercury perihelion advance anomaly. This anomaly turned out to be the first relativistic-gravity effect observed. During the 141 years to 2000, the precisions of laboratory and space experiments, and astrophysical and cosmological observations on relativistic gravity have been improved by 3 orders of magnitude. In 1999, we envisaged a 3-6 order improvement in the next 30 years in all directions of tests of relativistic gravity. In 2000, the interferometric gravitational wave detectors began their runs to accumulate data. In 2003, the measurement of relativistic Shapiro time-delay of the Cassini spacecraft determined the relativistic-gravity parameter γ to be 1.000021 ± 0.000023 of general relativity --- a 1.5-order improvement. In October 2004, Ciufolini and Pavlis reported a measurement of the Lense-Thirring effect on the LAGEOS and LAGEOS2 satellites to be 0.99 ± 0.10 of the value predicted by general relativity. In April 2004, Gravity Probe B (Stanford relativity gyroscope experiment to measure the Lense-Thirring effect to 1 %) was launched and has been accumulating science data for more than 170 days now. μSCOPE (MICROSCOPE: MICRO-Satellite à trainée Compensée pour l'Observation du Principle d'Équivalence) is on its way for a 2007 launch to test Galileo equivalence principle to $10^{-15}$. LISA Pathfinder (SMART2), the technological demonstrator for the LISA (Laser Interferometer Space Antenna) mission is well on its way for a 2008 launch. STEP (Satellite Test of Equivalence Principle), and ASTROD (Astrodynamical Space Test of Relativity using Optical Devices) are in the good planning stage. Various astrophysical tests and cosmological tests of relativistic gravity will reach precision and ultra-precision stages. Clock tests and atomic interferometry tests of relativistic gravity will reach an ever-increasing precision. These will give revived interest and development both in experimental and theoretical aspects of gravity, and may lead to answers to some profound questions of gravity and the cosmos.

## 1. Introduction

A dimensionless parameter $\xi(\mathbf{x}, t)$ characterizing the strength of gravity at a spacetime point $\wp$ [with coordinates $(\mathbf{x}, t)$] due to a gravitating source is the ratio of the negative of the potential energy, $mU$ (due to this source), to the inertial mass-energy $mc^2$ of a test body at $\wp$, i.e.,

$$\xi(\mathbf{x}, t) = U(\mathbf{x}, t) / c^2. \tag{1}$$

Here $U(\mathbf{x}, t)$ is the gravitational potential.

For a point source with mass $M$ in Newtonian gravity,

$$\xi(\mathbf{x}, t) = GM / Rc^2, \tag{2}$$

where $R$ is the distance to the source. For a nearly Newtonian system, we can use Newtonian potential for $U$. The strength of gravity for various configurations is tabulated in Table 1.

Table 1.  The strength of gravity for various configurations.

| Source | Field Position | Strength of Gravity $\xi$ |
|---|---|---|
| Sun | Solar Surface | $2.1 \times 10^{-6}$ |
| Sun | Mercury Orbit | $2.5 \times 10^{-7}$ |
| Sun | Earth Orbit | $1.0 \times 10^{-8}$ |
| Sun | Jupiter Orbit | $1.9 \times 10^{-9}$ |
| Earth | Earth Surface | $0.7 \times 10^{-9}$ |
| Earth | Moon's Orbit | $1.2 \times 10^{-11}$ |
| Galaxy | Solar System | $10^{-5} - 10^{-6}$ |
| Significant Part of Observed Universe | Our Galaxy | $1 - 10^{-2}$ |

The development of gravity theory stems from experiments. Newton's theory of gravity [1] is empirically based on Kepler's laws [2] (which are based on Brahe's observations) and Galileo's law of free-falls [3] (which is based on Galileo's experiment of motions on inclined planes and with free-all trajectories). Toward the middle of nineteenth century, astronomical observations accumulated a precision which enabled Le Verrier [4] to discover the mercury perihelion advance anomaly in 1859. This anomaly is the first relativistic-gravity effect observed. Michelson-Morley experiment [5], via various developments [6], prompted the final establishment of the special relativity theory in 1905 [7, 8]. Motivation for putting electromagnetism and gravity into the same theoretical framework [7], the precision of Eötvös experiment [9] on the equivalence, the formulation of Einstein equivalence principle [10] together with the perihelion advance anomaly led to the road for general relativity theory [11] in 1915.

In 1919, the observation of gravitational deflection of light passing near the Sun during a solar eclipse [12] made general relativity famous and popular. The perihelion advance of Mercury, the gravitational deflection of light passing near the Sun together with the establishment of the gravitational redshift constitute three classical tests of general relativity. With the development of technology and advent of space era, Shapiro [13] proposed a fourth test --- time delay of radar echoes in gravitational field. This, together with the more precise test of the equivalence principle [14] and Pound-Rebka redshift experiment [15] in 1960's, marked the beginning of a new era for testing relativistic gravity. Since the beginning of this era, we have seen 3-4 order improvements for old tests together with many new tests. The technological development is ripe that we are now in a position to discern 3-5 order further improvements in testing relativistic gravity in the coming 25 years (2005-2030). This will enable us to test the second order relativistic-gravity effects. A road map of experimental progress in gravity together with its theoretical implication is shown in Table 2.

Table 2. A road map (Highlights) for gravity. D denotes dynamical effect; EP denotes equivalence principle effect.

| Experiment | Precision † | Theory |
|---|---|---|
| Brahe's observations (1584-1600) (2'-5' accuracy) | $10^{-3}$ (D) | Kepler's laws |
| Galileo's experiment on inclined planes (1592) | $5 \times 10^{-3}$ (EP) $5 \times 10^{-3}$ (D) | (i) Galileo's EP (ii) The motion with constant force has constant acceleration. |
| Newton's pendulum experiment (1687 [1]) | $10^{-3}$ (EP) | § |
| Observation of celestial-body motions of the solar system (~1687) | $10^{-3}$ (EP) $10^{-3}$ - $10^{-4}$ (D) | Newton's inverse square law |
| Anomalous advance of Mercury's perihelion based on 397 meridian and 14 transit observations (1859 [4]) | $10^{-8}$ (D) | * |
| Anomalous advance of Mercury's perihelion based on transit observations from 1677 to 1881 (1882 [16]) | $10^{-9}$ (D) | * |
| Michelson-Morley Experiment (1887 [5]) | $10^{-9}$ | Special relativity * |
| Eötvös experiment (1889 [9]) | $10^{-8}$ (EP) | §, * |
| Light deflection experiment (1919 [12]) | $5 \times 10^{-7}$ (D) | # |
| Roll-Krotkov-Dicke experiment (1964 [14]) | $10^{-11}$ (EP) | § |
| Binary pulsar observation (1979 [17]) | $10^{-13}$ (D) | ¶ |
| Solar system test (1970-1999) | $10^{-10}$ - $10^{-11}$ (D) | ‖ |
| Supernova Cosmology Experiment (1998-1999 [18-21]) | $10^{-2}$ (D) | General relativity with cosmological constant |
| Cassini time-delay experiment [56] | $10^{-11}$ - $10^{-12}$ (D) | # |
| LAGEOS gravity experiment [67] | $10^{-11}$ (D) | %, # |
| GP-B Experiment [68] | $10^{-12}$ (D) | % |
| µSCOPE Experiment [109] | $10^{-15}$ (EP) | ¦ |
| STEP (2010-2030) | $10^{-17}$ - $10^{-19}$ (EP) | ¦ |
| Bepi-Colombo, GAIA, ASTROD (2010-2030) | $10^{-14}$ - $10^{-17}$ (D) | ‡ |

† In terms of dominant observable effects

§ Confirmation of Galileo's equivalence principle

* Leading to general relativity

# Confirming the prediction of general relativity

¶ Confirming the quadrupole radiation formula of general relativity

‖ In terms of relativistic parameters, the precision is $10^{-3}$

% Testing frame dragging of general relativity

¦ Testing Galileo's equivalence principle

‡ In terms of relativistic parameters, the precision will be $10^{-5}$ - $10^{-9}$

This review is a five-year update from a previous review article (W.-T. Ni, "Empirical tests of the relativistic gravity: the past, the present and the future", in *Recent advances and cross-century outlooks in physics: interplay between theory and experiment: proceedings of the Conference held on March 18-20, 1999 in Atlanta, Georgia,* editors, Pisin Chen, and Cheuk-Yin Wong, [Singapore: World Scientific, 2000] ; and pp. 1-19 in *Gravitation and Astrophysics,* editors, Liu L, Luo J, Li X-Z and Hsu J-P [Singapore: World Scientific, 2000] ).

In section 2, we review the Mercury's perihelion advance and events leading to general relativity. In section 3, we discuss the classical tests. In section 4, we review precision measurement tests and the foundations of relativistic gravity. In section 5, we review solar system tests since the revival (1960). In sections 6 and 7, we discuss astrophysical tests and cosmological tests respectively. In section 8, we discuss gravitational-wave observations in relation to testing relativistic gravity. In section 9, we discuss next generation experiments in progress, planned and proposed. In section 10, we give an outlook. In the appendix, we discuss empirical tests associated with Eddittington-Robertson formalism.

## 2. Mercury's Perihelion Advance and Events Leading to General Relativity

In 1781, Herschel discovered the planet Uranus. Over years, Uranus persistently wandered away from its expected Newtonian path. In 1834, Hussey suggested that the deviation is due to perturbation of an undiscovered planet. In 1846, Le Verrier predicted the position of this new planet. On September 25, 1846, Galle and d'Arrest found the new planet, Neptune, within one degree of arc of Le Verrier's calculation. This symbolized the great achievement of Newton's theory. [22]

With the discovery of Neptune, Newton's theory of gravitation was at its peak. As the orbit determination of Mercury reached $10^{-8}$, relativistic effect of gravity showed up. In 1859, Le Verrier discovered the anomalous perihelion advance of Mercury [4].

In 1840, Arago suggested to Le Verrier to work on the subject of Mercury's motion. Le Verrier published a provisional theory in 1843. It was tested at the 1848 transit of Mercury and there was not close agreement. As to the cause, Le Verrier [23] wrote "Unfortunately, the consequences of the principle of gravitation have not been deduced in many particulars with a sufficient rigour: we will not be able to decide, when faced with a disagreement between observation and theory, whether this results completely from analytical errors or whether it is due in part to the imperfection of our knowledge of celestial physics." [24]

In 1859, Le Verrier [4] published a more sophisticated theory of Mercury's motion. This theory was sufficently rigorous for any disagreement with observation to be taken quite confidently as indicating a new scientific fact. In this paper, he used two sets of observations --- a series of 397 meridian observations of Mercury taken at the Paris Observatory between 1801 and 1842, and a set of observations of 14 transits of Mercury. The transit data are more precise and the uncertainty is of the order of 1". The calculated planetary perturbations of Mercury is listed in Table 3. In addition to these perturbations, there is a 5025"/century general precession in the observational data due to the precession of equinox. The fit of observational data with theoretical calculations has discrepancies. These discrepancies turned out to be due to relativistic-gravity effects. Le Verrier attributed these discrepancies to an additional 38" per century anomalous advance in the perihelion of Mercury. [24]

Newcomb [16] in 1882, with improved calculations and data set, obtained 42".95 per century anomalous perihelion advance of Mercury. The value more recently was (42".98 ± 0.04)/century [25].

In the last half of the 19th century, efforts to account for the anomalous perihelion advance of Mercury went into two general directions: (i) searching for the planet Vulcan, intra-Mercurial matter and the like; (ii) modification of the gravitation law. Both kinds of efforts were not successful. For modification of the gravitational law, Clairaut's hypothesis, Hall's hypothesis and velocity-dependent force laws were considered. The successful solution awaited the development of general relativity.

In 1887, the result of Michelson-Morley experiment posed a serious problem to Newtonian mechanics. A series of developments [6] led to Poincaré's adding to the five classical principles of Physics the Principle of Relativity [26, 27] in 1904 --- "*The laws of physical phenomena must be the same for a fixed observer and for an observer in rectilinear and uniform motion so that we have no possibility of perceiving whether or not we are dragged in such a motion*", and to the seminal works of Poincaré [7] and Einstein [8] in 1905. In [7], Poincaré attempted to develop a relativistic theory of gravity and mentioned gravitational-wave propagating with the speed of light based on Lorentz invariance.

A crucial milestone toward a viable relativistic theory of gravity was established when Einstein proposed his equivalence principle in 1907 [10]. This principle had a firm empirical basis due to the precision experiment of Eötvös [9]. In this same paper, Einstein predicted gravitational redshift. With Special Relativity and the Einstein Equivalence Principle, geometrization of classical physics in the large came naturally in the setting of pseudo-Riemannian manifold. In the years up to 1915 were full of debates and arguments between a number of physicists (Abraham, Einstein, Nordstrom,....) concerned with developing a new relativistic gravitational theory [28]. In 1915, Einstein's General Relativity was proposed [11, 29] and the anomalous perihelion advance of Mercury was explained.

Table 3. Planetary perturbations of the perihelion of Mercury [4]

| | |
|---|---|
| Venus | 280".6/century |
| Earth | 83".6/century |
| Mars | 2".6/century |
| Jupiter | 152".6/century |
| Saturn | 7".2/century |
| Uranus | 0".1/century |
| Total | 526".7/century |

### 3. Classical Tests

The perihelion advance anomaly of Mercury, the deflection of light passing the limb of the Sun and the gravitational redshift are the three classical tests of relativistic gravity. Using EEP, Einstein [30] derived the deflection of light passing the limb of the Sun in 1911. This agrees with the deflection of light derived by using particle model of light in the late 18th century. Before 1915, observations on light deflection were not successful due to war and weather. Einstein's general relativity doubled the prediction of the deflection of light (1".75). The 1919 British solar eclipse expeditions reported reasonably good agreement with the prediction of Einstein's relativity. Before 1960, there were several such observations. The accuracy of these observations was not better than 10 - 20%.

After Einstein [10] proposed the gravitational redshift, Freundlich [31] started the long effort to disentangle the gravitational redshift of solar and other stellar spectral lines from other causes. Over the next five decades, astronomers did not agree on whether there is gravitational redshift empirically [32]. This question is finally settled and gravitational redshift confirmed by Pound and Rebka [15] using Mössbauer effect. The improved result of Pound and Snider [15] confirmed the redshift prediction to 1 % accuracy.

## 4. Precision Measurement and Foundations of Relativistic Gravity

The foundation of relativistic gravity at present rests on the equivalence of local physics everywhere in spacetime. This equivalence is called the Einstein equivalence principle (EEP) [10]. Its validity guarantees the universal implementation of metrology and standards. Precision metrology or measurement, in turn, test its validity. Possible violations will give clues to the origin of gravity.

The most tested part of equivalence is the Galileo equivalence principle (the universality of free-all). In the study of the theoretical relations between the Galileo equivalence principle and the Einstein equivalence principle, we [33, 34] proposed the $\chi$-g framework summarized in the following interaction Lagrangian density

$$L_I = - (1/(16\pi))\chi^{ijkl} F_{ij} F_{kl} - A_k{}_i^k (-g)^{(1/2)} - \Sigma_I m_I (ds_I)/(dt) \, \delta(\mathbf{x}-\mathbf{x}_I), \qquad (3)$$

where $\chi^{ijkl} = \chi^{klij} = - \chi^{klji}$ is a tensor density of the gravitational fields (e.g., $g_{ij}$, $\varphi$, etc.) and $j^k$, $F_{ij} \equiv A_{j,i} - A_{i,j}$ have the usual meaning. The gravitation constitutive tensor density $\chi^{ijkl}$ dictates the behavior of electromagnetism in a gravitational field and has 21 independent components in general. For a metric theory (when EEP holds), $\chi^{ijkl}$ is determined completely by the metric $g^{ij}$ and equals $(-g)^{1/2} [(1/2) g^{ik} g^{jl} - (1/2) g^{il} g^{kj}]$. Here we use this framework to look into the foundation of relativistic gravity empirically.

The condition for no birefringence (no splitting, no retardation) for electromagnetic wave propagation in all directions in the weak field limit gives ten constraints on the $\chi$'s. With these ten constraints, $\chi$ can be written in the following form

$$\chi^{ijkl} = (-H)^{1/2}[(1/2)H^{ik} H^{jl} - (1/2)H^{il} H^{kj}]\psi + \varphi e^{ijkl}, \qquad (4)$$

where $H = \det(H_{ij})$ is a metric which generates the light cone for electromagnetic propagation, and $e^{ijkl}$ is the completely antisymmetric symbol with $e^{0123} = 1$ [35-37]. Recently, Lämmerzahl and Hehl have shown that this non-birefringence guarantees, without approximation, Riemannian light cone, i.e., Eq. (4) [38].

Eq. (4) is verified empirically to high accuracy from pulsar observations and from polarization measurements of extragalactic radio sources and will be discussed in §6 on the astrophysical tests. Let us now look into the empirical constraints for $H^{ij}$ and $\varphi$. In Eq. (3), $ds$ is the line element determined from the metric $g_{ij}$. From Eq. (4), the gravitational coupling to electromagnetism is determined by the metric $H_{ij}$ and two scalar fields $\varphi$ and $\psi$. If $H_{ij}$ is not proportional to $g_{ij}$, then the hyperfine levels of the lithium atom, the beryllium atom, the mercury atom and other atoms will have additional shifts. But this is not observed to high accuracy in Hughes-Drever experiments [39]. Therefore $H_{ij}$ is proportional to $g_{ij}$ to a certain accuracy. Since a change of $H^{ij}$ to $\lambda H^{ij}$ does not affect $\chi^{ijkl}$ in Eq. (4), we can define $H_{11} = g_{11}$ to remove this scale freedom. [35, 40]

In Hughes-Drever experiments [39] $\Delta m/m \leq 0.5 \times 10^{-28}$ or $\Delta m/m_{e.m.} \leq 0.3 \times 10^{-24}$ where $m_{e.m.}$ is the electromagnetic binding energy. Using Eq. (4) in Eq. (3), we have three kinds of contributions to $\Delta m/m_{e.m.}$. These three kinds are of the order of (i) $(H_{\mu\nu} - g_{\mu\nu})$, (ii) $(H_{0\mu} - g_{0\mu})v$, and (iii) $(H_{00} - g_{00})v^2$ respectively [35, 40]. Here the Greek indices $\mu$, $\nu$ denote space indices. Considering the motion of laboratories from earth rotation, in the solar system and in our galaxy, we can set limits on various components of $(H_{ij} - g_{ij})$ from Hughes-Drever experiments as follows:

$$| H_{\mu\nu} - g_{\mu\nu} | / U \leq 10^{-18}$$
$$| H_{0\mu} - g_{0\mu} | / U \leq 10^{-13} - 10^{-14},$$
$$| H_{00} - g_{00} | / U \leq 10^{-10}. \tag{5}$$

where U ($\sim 10^{-6}$) is the galactical gravitational potential.

Eötvös-Dicke experiments [9, 14, 41-43] are performed on unpolarized test bodies; the latest such experiments [43] reach a precision of $3 \times 10^{-13}$. In essence, these experiments show that unpolarized electric and magnetic energies follow the same trajectories as other forms of energy to certain accuracy. The constraints on Eq. (4) are

$$| 1 - \psi | / U < 10^{-10} \tag{6}$$

and

$$| H_{00} - g_{00} | / U < 10^{-6} \tag{7}$$

where U is the solar gravitational potential at the earth.

In 1976, Vessot and Levine [44] used an atomic hydrogen maser clock in a space probe to test and confirm the metric gravitational redshift to an accuracy of $1.4 \times 10^{-4}$ [45]. The space probe attained an altitude of 10,000 km above the earth's surface. With Eq. (6), the constraint on Eq. (4) is

$$| H_{00} - g_{00} | / U \leq 1.4 \times 10^{-4}. \tag{8}$$

Thus, we see that for the constraint on $| H_{00} - g_{00} | / U$, Hughes-Drever experiments give the most stringent limit. However, STEP mission concept [46] proposes to improve the WEP experiment by five orders of magnitude. This will again lead in precision in determining $H_{00}$.

The theory (3) with $\chi^{ijkl}$ given by

$$\chi^{ijk} = (-g)^{1/2} [(1/2) \, g^{ik} \, g^{jl} - (1/2) \, g^{il} \, g^{kj} + \varphi \, \varepsilon^{ijkl}] \,, \tag{9}$$

where $\varphi$ is a scalar or pseudoscalar function of the gravitational field and $\varepsilon^{ijkl} = (-g)^{-1/2} e^{ijkl}$ is studied in [47] and [48]. In (3), particles considered have charges but no spin. To include spin-1/2 particles, we can add the Lagrangian for Dirac particles. Experimental tests of the equivalence principle for polarized-bodies are reviewed in [49].

To include QCD and other gauge interactions, we have generalized the $\chi$-$g$ framework [50]. Now we are working on a more comprehensive generalization to include a framework to test special relativity, and a framework to test the gravitational interactions of scalar particles and particles with spins together with gauge fields.

## 5. Solar System Tests

For last forty years, we have seen great advances in the dynamical testing of relativistic gravity. This is largely due to interplanetary radio ranging and lunar laser ranging. Interplanetary radio ranging and tracking provided more stimuli and progresses at first. However with improved accuracy of 2 cm from 20-30 cm and long-accumulation of observation data, lunar laser ranging reaches similar accuracy in determining relativistic parameters as compared to interplanetary radio ranging. Table 4 gives such a comparison.

Table 4. Relativity-parameter determination from interplanetary radio ranging and from lunar laser ranging.

| Parameter | Meaning | Value from Solar System Determinations | Value from Lunar Laser Ranging |
|---|---|---|---|
| $\beta$ | PPN [51] Nonlinear Gravity | $1.000\pm0.003$ [25] (perihelion shift with $J_2$ (Sun)=$10^{-7}$ assumed) <br> $0.9990\pm0.0012$ [52] (Solar-System Tests with $J_2$ (Sun)=$(2.3\pm5.2)\times10^{-7}$ fitted) <br> $1.0000\pm0.0001$ [53] (EPM2004 fitting) | $1.003\pm0.005$ [54] <br><br> $1.00012\pm0.0011$ [55, 56] |
| $\gamma$ | PPN Space Curvature | $1.000\pm0.002$ [25] (Viking ranging time delay) <br> $0.9985\pm0.0021$ [52] (Solar-System Tests) <br> $1.000021\pm0.000023$ [56](Cassini S/C Ranging) <br> $0.9999\pm0.0001$ [53] (EPM2004 fitting) | $1.000\pm0.005$ [54] |
| $K_{gp}$ | Geodetic Precession | | $0.997\pm0.007$ [54] <br> $0.9981\pm0.0064$ [55] |
| E | Strong Equivalence Principle | | $(3.2\pm4.6)\times10^{-13}$ [54] <br> $(-2.0\pm2.0)\times10^{-13}$ [55, 43] |
| $\dot{G}/G$ | Temporal Change in G | $(2\pm4)\times10^{-12}$/yr [57] (Viking Lander Ranging) <br> $\pm10\times10^{-12}$/yr [58] (Viking Lander Ranging) <br> $\pm2.0\times10^{-12}$/yr [59](Mercury & Venus Ranging) <br> $\pm(1.1-1.8)\times10^{-12}$/yr [60] (Solar-System Tests) | $(1\pm8)\times10^{-12}$/yr [54] <br> $(0.4\pm0.9)\times10^{-12}$/yr [55] |

In the last column of Table 4, the values come from two references [54] and [55]. In [55], Williams *et al.* used a total of 15 553 LLR normal-point data in the period of March 1970 to April 2004 from Observatoire de la Côte d'Azur, McDonald Observatory and Haleakala Observatory in their determination. Each normal point comprises from 3 to about 100 photons. The weighted rms scatter after their fits for the last ten years of ranges is about 2 cm (about $5 \times 10^{-11}$ of range).

In 2003, Bertotti, Iess and Tortora [56] reported a measurement of the frequency shift of radio photons due to relativistic Shapiro time-delay effect from the Cassini spacecraft as they passed near the Sun during the June 2002 solar conjunction. From this measurement, they determined $\gamma$ to be $1.000021 \pm 0.000023$.

With the advent of VLBI (Very Long Baseline Interferometry) at radio wavelengths, the gravitational deflection of radio waves by the Sun from astrophysical radio sources has been observed and accuracy of observation had been improved to $1.7 \times 10^{-3}$ for $\gamma$ [61, 62] in 1995. Recent analysis using VLBI data from 1979-1999 improved this result by about four times to $0.99983 \pm 0.00045$ [63]. Fomalont and Kopeikin [64] measured the effect of retardation of gravity by the field of moving Jupiter via VLBI observation of light bending from a quasar.

The solar-system measurements have made possible the creation of high-accuracy planetary and lunar ephemerides. Two most complete series of ephemerides are the numerical DE ephemerides of JPL [65] and the EPM ephemerides of the Institute of Applied Astromomy [53]. They are of the same level of accuracy and can be used to fit experiments/observations and to determine astronomical constants. Krasinsky and Brumberg [66] used these two series of ephemerides to analyze the major planet motions and the AU (Astronomical Unit); Pitjeva [53] have recently used the EPM framework to determine the AU and obtain 1 AU = 149 597 870 696.0 m. The JPL DE410 determination of this number is 1 AU = 149 597 870 697.4 m. The difference of 1.4 m represents the realistic error in the determination of the AU. Pitjeva's [53] determination of β and γ is obtained simultaneously with estimations for the solar oblateness and the possible variability of the gravitational constant.

In 1918, Lense and Thirring predicted that the rotation of a body like Earth will drag the local inertial frames of reference around it in general relativity. In 2004, Ciufolini and Pavlis [67] reported a measurement of this Lense-Thirring effect on the two Earth satellites, LAGEOS and LAGEOS2; it is 0.99 ± 0.10 of the value predicted by general relativity. In the same year, Gravity Probe B (a space mission to test general relativity using cryogenic gyroscopes in orbit) was launched in April and aims at measurement of Lense-Thirring effect to about 1 % [68].

With the Hipparcos mission, very accurate measurements of star positions at various elongations from the Sun were accumulated. Most of the measurements were at elongations greater than 47° from the Sun. At these angles, the relativistic light deflections are typically a few mas; it is 4.07 mas according to general relativity at right angles to the solar direction for an observer at 1 AU from the Sun. In the Hipparcos measurements, each abscissa on a reference great-circle has a typical precision of 3 mas for a star with 8-9 mag. There are about 3.5 million abscissae generated, and the precision in angle or similar parameter determination is in the range. Frœschlé, Mignard and Arenou [69] analyzed these Hipparcos data and determined the light deflection parameter γ to be 0.997 ± 0.003. This result demonstrated the power of precision optical astrometry.

## 6. Astrophysical Tests

In the early days, astronomical observations of the solar system provided the basis for developing gravitation theories. With increasing precise observations, astrophysics and cosmology are increasingly more important for such developments. Precise timing of pulsars provides:

(i) confirmation of quadrupole radiation formula for gravitational radiation [17],

(ii) additional testing ground for Post-Newtonian Parameters [17],

(iii) test of nonbirefringence of propagation of electromagnetic wave in a gravitational field, and

(iv) upper limit of background gravitational-wave radiation [70-73].

We refer (i), (ii) and (iv) to references cited. Here, we discuss (iii).

With the null-birefringence observations of pulsar pulses and micropulses before 1980, the relations (4) for testing EEP are empirically verified to $10^{-14} - 10^{-16}$ [35-37]. With the present pulsar observations, these limits would be improved; a detailed such analysis is in [74]. Analyzing the data from polarization measurements of extragalactic radio sources, Haugan and Kauffmann [75] inferred that the resolution for null-birefringence is 0.02 cycle at 5 GHz. This corresponds to a time resolution of $4 \times 10^{-12}$ s and gives much better constraints. With a detailed analysis and more extragalactic radio observations, (4) would be tested down to $10^{-28}$-$10^{-29}$ at cosmological distances. In 2002, Kostelecky and Mews [76] used polarization measurements of light from cosmologically distant astrophysical sources to yield stringent constraints down to $2 \times 10^{-32}$. The electromagnetic propagation in Moffat's nonsymmetric gravitational theory fits the χ-g framework. Krisher [77], and Haugan and Kauffmann [75] have used the pulsar data and extragalactic radio observations to constrain it.
.

## 7. Cosmological Tests

In an attempt to find a static cosmology, Einstein add a cosmological constant $\Lambda$ to his equation. The term containing $\Lambda$ can be interpreted as a modification of Einstein's equation or it can be just interpreted as vacuum stress-energy.

Although Einstein considered the proposal of this term his biggest blunder in his life, the value of $\Lambda$ needs to be determined using cosmological observations.

Recent evidence suggests that Type Ia supernovae (SNeIa) can be used as precise cosmological distance indicators [78]. Early results with these SNeIa observations imply that there is not enough gravitating matter to close the universe [18, 19] and that currently the expansion of the Universe is accelerating [20, 21], indicating $\Lambda$-density (cosmological term, dark energy or quintessence) is larger than the ordinary-matter density. More supernovae observations together with more precise cosmic background anisotropy measurements will be important in testing and determining the gravitational equation in the cosmological context.

In section 4, we mentioned a nonmetric theory [33, 34, 47] (including electromagnetic interaction of axion theories) in discussing the foundations of relativistic gravity. Theories with spontaneous direction [79] also have such an electromagnetic interaction. The effect of φ [in (9)] in this theory is to change the phase of two different circular polarizations in gravitation field and gives polarization rotation for linearly polarized light [47, 79, 80]. Using polarization observations of radio galaxies, Carroll, Field and Jackiw [79, 80] put a limit of 0.1 on $\Delta\varphi$ over cosmological distances. Using a different analysis of polarization observation of radio galaxies, Nodland and Ralston [81] found indication of anisotropy in electromagnetic propagation over cosmological distances with a birefringence scale of order $10^{25}$ m (i.e., about 0.1 - 0.2 Hubble distance). This gave $\Delta\varphi \sim 5$ - 10 over Hubble distance). Later analyses [82-86] did not confirm this result and put a limit of $\Delta\varphi \leq 1$ over cosmological distance scale.

The natural coupling strength φ is of order 1. However, the isotropy of our observable universe to $10^{-5}$ may leads to a change $\Delta\varphi$ of φ over cosmological distance scale $10^{-5}$ smaller [47, 87]. Hence, observations to test and measure $\Delta\varphi$ to $10^{-6}$ are significant and they are promising. In 2002, DASY microwave interferometer observed the polarization of the cosmic background. With the axial interaction (9), the polarization anisotropy is shifted relative to the temperature anisotropy. In 2003, WMAP (Wilkenson Microwave Anisotropy Probe) [88] found that the polarization and temperature are correlated. This gives a constraint of $10^{-1}$ of $\Delta\varphi$ [89]. Planck Surveyor [90] will be launched in 2007 with better polarization-temperature measurement and will give a sensitivity to $\Delta\varphi$ of $10^{-2}$-$10^{-3}$. A dedicated future experiment on cosmic microwave background radiation polarization will reach $10^{-5}$-$10^{-6}$ $\Delta\varphi$-sensitivity. This is very significant as a positive result may indicate that our patch of inflationary universe has a 'spontaneous polarization' in fundamental law of electromagnetic propagation influenced by neighboring patches and we can 'observe' neighboring patches through a determination of this fundamental physical law; if a negative result turns out at this level, it may give a good constraint on superstring theories as axions are natural to superstring theories.

In section 4, we mentioned Hughes-Drever experiments [39] to test the spatial anisotropy. Here we mention test of cosmic spatial anisotropy using polarized electrons. Following Phillips' pioneer work [91], we [92] and Berglund *et al.* [93] improved the sensitivity. In 2000, we used a rotatable torsion balance carrying a transversely spin-polarized ferrimagnetic $Dy_6Fe_{23}$ mass to test the cosmic spatial anisotropy, and have achieved an order-of-magnitude improvement [94, 95] over previous experiments; the anomalous transverse energy splitting of spin states of electron due to spatial anisotropy is constrained to be less than $6 \times 10^{-29}$ GeV. Eöt-Wash group also improved on their experiment and gave a constraint of $1.2 \times 10^{-28}$ GeV [96].

## 8. Gravitational-Wave Observations

The importance of gravitational-wave detection is twofold: (i) as probes to fundamental physics and cosmology, especially black hole physics and early cosmology, and (ii) as tools in astronomy and astrophysics to study compact objects and to count them. We follow [97] to extend the conventional classification of gravitational-wave frequency bands [98] into the ranges:

(i) High-frequency band (1-10 kHz): This is the frequency band that ground gravitational-wave detectors are most sensitive to.

(ii) Low-frequency band (100 nHz - 1 Hz): This is the frequency band that space gravitational-wave experiments are most sensitive to.

(iii) Very-low-frequency band (300 pHz-100 nHz): This is the frequency band that the pulsar timing experiments are most sensitive to.

(iv) Extremely-low-frequency band (1 aHz - 10 fHz): This is the frequency band that the cosmic microwave anisotropy and polarization experiments are most sensitive to.

The cryogenic resonant bar detectors have already reached a strain sensitivity of $(10^{-21})/(Hz)^{1/2}$ in the kHz region. Five such detectors --- ALLEGRO, AURIGA, EXPLORER, NAUTILUS and NIOBE --- have been on the air, forming a network with their bar axis quasi-parallel in a continuous search for bursts. TAMA (300 m armlength) interferometer started accumulating data in 2000. GEO, and kilometer size laser-interferometric gravitational-wave detectors --- LIGO and VIRGO --- took runs and started to accumulate data also with strain sensitivity goal aimed at $10^{-23}/(Hz)^{1/2}$ in the frequency around 100 Hz. Various limits on the gravitation-wave strains for different sources become significant. For example, analysis of data collected during the second LIGO science run set strain upper limits as low as a few times $10^{-24}$ for some pulsar sources; these translate into limits on the equatorial ellipticities of the pulsars, which are smaller than $10^{-5}$ for the four closest pulsars [99].

Space interferometer (LISA [100, 101], ASTROD [102, 103]) for gravitational-wave detection hold the most promise. LISA (Laser Interferometer Space Antenna) [100] is aimed at detection of low-frequency ($10^{-4}$ to 1 Hz) gravitational waves with a strain sensitivity of $4 \times 10^{-21}/(Hz)^{1/2}$ at 1 mHz. There are abundant sources for LISA and ASTROD: galactic binaries (neutron stars, white dwarfs, etc.). Extra-galactic targets include supermassive black hole binaries, supermassive black hole formation, and cosmic background gravitational waves. A date of LISA launch is hoped for 2013.

For the very-low-frequency band and for the extremely-low-frequency band, it is more convenient to express the sensitivity in terms of energy density per logarithmic frequency interval divided by the cosmic closure density $\rho_c$ for a cosmic background of gravitational waves, i.e., $\Omega_g(f)(=(f/\rho_c)d\rho_g(f)/df)$.

The upper limits from pulsar timing observations on a gravitational wave background are about $\Omega_g \leq 10^{-7}$ in the frequency range 4-40 nHz [70], and $\Omega_g \leq 4 \times 10^{-9}$ at $6 \times 10^{-8}$ Hz [72]. More pulsar observations with extended periods of time will improve the limits by two orders of magnitude in the lifetime of present ground and space gravitational-wave-detector projects. The COBE microwave-background quadrupole anisotropy measurement [104, 105] gives a limit $\Omega_g$ (1 aHz) $\sim 10^{-9}$ on the extremely-low-frequency gravitational-wave background [106, 107]. Ground and balloon experiments probe smaller-angle anisotropies and, hence, higher-frequency background. WMAP [108] and Planck Surveyor [90] space missions can probe anisotropies with *l* up to 2000 and with higher sensitivity.

## 9. The Next 25 Years (2005-2030)

During last 146 years, the precisions of laboratory and space experiments, and astrophysical and cosmological observations on relativistic gravity have improved by 3-4 orders of magnitude. In 2000, the ground long interferometric gravitational wave detectors began their runs to accumulate data. In April 2004, Gravity Probe B (Stanford relativity gyroscope experiment to measure the Lense-Thirring effect to 1 %) [68] was launched and has been accumulating science data for more than 170 days now. µSCOPE (MICROSCOPE: MICRO-Satellite à trainée Compensée pour l'Observation du Principe d'Équivalence) [109] is on its way for a 2007 launch to test Galileo equivalence principle to $10^{-15}$. LISA Pathfinder (SMART2), the technological demonstrator for the LISA (Laser Interferometer Space Antenna) mission is well on its way for a 2008 launch. STEP (Satellite Test of Equivalence Principle) [46], and ASTROD (Astrodynamical Space Test of Relativity using Optical Devices) [102, 103] are in the good planning stage. Various astrophysical tests and cosmological tests of relativistic gravity will reach precision and ultra-precision stages. Clock tests and atomic interferometry tests of relativistic gravity will reach an ever-increasing precision. In the next 25 years, we envisage a 3-5 order improvement in all directions of tests of relativistic gravity. These will give revived interest and development both in experimental and theoretical aspects of gravity, and may lead to answers to some profound questions of gravity and the cosmos.

In this section, we illustrate this expectation by looking into various ongoing / proposed experiments related to the determination of the PPN space curvature parameter γ. Some motivations for determining γ precisely to $10^{-5} - 10^{-9}$ are given in [110, 111]. Table 5 lists the aimed accuracy of such experiments.

GP-B is an ongoing experiment [68] using quartz gyro at low-temperature to measure the Lense-Thirring precession and the geodetic precession. The geodetic precession gives a measure of γ.

Bepi-Colombo [112] is planned for a launch in 2013 to Mercury. A simulation predicts that the determination of γ can reach $2 \times 10^{-6}$ [113].

GAIA (Global Astrometric Interferometer for Astrophysics) [114] is an astrometric mission concept aiming at the broadest possible astrophysical exploitation of optical interferometry using a modest baseline length (~3m). GAIA is planned to be launched in 2013. At the present study, GAIA aims at limit magnitude 21, with survey completeness to visual magnitude 19-20, and proposes to measure the angular positions of 35 million objects (to visual magnitude V=15) to 10 $\mu$as accuracy and those of 1.3 billion objects (to V=20) to 0.2 mas accuracy. The observing accuracy of V=10 objects is aimed at 4 $\mu$as. To increase the weight of measuring the relativistic light deflection parameter $\gamma$, GAIA is planned to do measurements at elongations greater than 35° (as compared to essentially 47° for Hipparcos) from the Sun. With all these, a simulation shows that GAIA could measure $\gamma$ to $1 \times 10^{-5} - 2 \times 10^{-7}$ accuracy [115].

Table 5. Aimed accuracy of PPN space parameter $\gamma$ for various ongoing / proposed experiments. The types of experiments (deflection, retardation or geodetic precession) are given in the parentheses.

| Ongoing / Proposed experiment | Aimed accuracy of $\gamma$ |
|---|---|
| GP-B [68] (geodetic precession) | $1 \times 10^{-5}$ |
| Bepi-Colombo [113] (retardation) | $2 \times 10^{-6}$ |
| GAIA [115] (deflection) | $1 \times 10^{-5} - 2 \times 10^{-7}$ |
| ASTROD I [116] (retardation) | $1 \times 10^{-7}$ |
| LATOR [117] | $1 \times 10^{-8}$ |
| ASTROD [118] | $1 \times 10^{-9}$ |

In the ranging experiments, the retardations (Shapiro time delays) of the electromagnetic waves are measured to give $\gamma$. In the astrometric experiments, the deflections of the electromagnetic waves are measured to give $\gamma$. These two kinds of experiments complement each other in determining $\gamma$. The ASTROD I (Single Spacecraft Astrodynamical Space Test of Relativity using Optical Devices) mission concept [116] is to use a drag-free spacecraft orbiting around the Sun using 2-way laser pulse ranging and laser-interferometric ranging between Earth and spacecraft to measure $\gamma$ and other relativistic parameters precisely. The $\gamma$ parameter can be separated from the study of the Shapiro delay variation. The uncertainty on the Shapiro delay measurement depends on the uncertainties introduced by the atmosphere, timing systems of the ground and space segments, and the drag-free noise. A simulation shows that an uncertainty of $10^{-7}$ on the determination of $\gamma$ is achievable.

LATOR (Laser Astrometric Test Of Relativity) [117] proposed to use laser interferometry between two micro-spacecraft in solar orbits, and a 100 m baseline multi-channel stellar optical interferometer placed on the ISS (International Space Station) to do spacecraft astrometry for a precise measurement of $\gamma$.

For ASTROD (Astrodynamical Space Test of Relativity) [102, 103, 118], 3 spacecraft, advanced drag-free systems, and mature laser interferometric ranging will be used and the resolution is subwavelength. The accuracy of measuring $\gamma$ and other parameters will depend on the stability of the lasers and/or clocks. An uncertainty of $1 \times 10^{-9}$ on the determination of $\gamma$ is achievable in the time frame of 2015-2025.

## 10. Outlook

Physics is an empirical science, so is gravitation. The road map for gravitation is clearly empirical. As precision is increased by orders of magnitude, we are in a position to explore deeper into the origin of gravitation. The current and coming generations are holding such promises.

## Acknowledgements

I would like to thank the National Natural Science Foundation (Grant No. 10475114), and the Foundation of Minor Planets of Purple Mountain Observatory for supporting this work.

## Appendix

Since many readers are more familiar with the parametrization given by Eddington (A. S. Eddington, *The Mathematical Theory of Relativity* [2$^{nd}$ ed., Cambridge University Press, 1924]) and Robertson (H. P. Robertson, p. 228 in *Space Age Astronomy*, ed. by A. J. Deutsch and W. H. Klemperer [Academic Press, New York, 1962]) in testing relativistic gravity, with the recommendation of J. P. Hsu (one of the editor of "*100 Years of Gravity and Accelerated Frames---The Deepest Insights of Einstein and Yang-Mills*" [Ed. J. P. Hsu and D. Fine, in press, World Scientific, 2005] in which this article is published simultaneously), I add this explanatory appendix.

The Eddington-Robertson parametrization of metric is

$$ds^2 = (1-2\alpha(GM/r)+2\beta(GM/r)^2+\ldots)dt^2 - (1+2\gamma(GM/r)+\ldots)\,(dr^2+r^2d\theta^2+r^2\sin^2\theta d\varphi^2),$$

where $\alpha$, $\beta$, $\gamma$ are the Eddington-Robertson relativistic parameters. For matter, $\alpha$ can be absorbed into G; the Newtonian limit then requires G to be the Newtonian gravitational constant. To test the Einstein Equivalence Principle, for electromagnetism, we could set $\alpha$, $\beta$ and $\gamma$ to $\alpha_{em}$, $\beta_{em}$ and $\gamma_{em}$, and the metric corresponds to $H_{ij}$ in section 4; for matter, we then have $\alpha$, $\beta$ and $\gamma$ equal to $\alpha_{matter}(=1)$, $\beta_{matter}$ and $\gamma_{matter}$, and the metric corresponds to $g_{ij}$. The constraints (5, 7, 8) become

$$2|\gamma_{em} - \gamma_{matter}| \leq 10^{-18},$$
$$2|\alpha_{em} - 1| \leq 10^{-10}, \tag{5'}$$

$$2|\alpha_{em} - 1| < 10^{-6}, \tag{7'}$$

$$2|\alpha_{em} - 1| \leq 1.4 \times 10^{-4}. \tag{8'}$$

For dynamical tests of $\beta$ and $\gamma$, Table 4 and Table 5 and the associated discussions in section 5 and section 9 apply.